\documentclass[10pt]{article}
\usepackage{graphicx}

\def\Title#1{\begin{center} {\Large #1 } \end{center}}
\def\Author#1{\begin{center}{ \sc #1} \end{center}}
\def\Address#1{\begin{center}{ \it #1} \end{center}}

\newcommand\pubblock{\rightline{\begin{tabular}{l} Proceedings of the Second Annual LHCP\\ \pubnumber\\
         \pubdate  \end{tabular}}}

\newenvironment{Abstract}{\begin{quotation} \begin{center} 
             \large ABSTRACT \end{center}\bigskip 
      \begin{center}\begin{large}}{\end{large}\end{center} \end{quotation}}

\newenvironment{Presented}{\begin{quotation} \begin{center} 
             PRESENTED AT\end{center}\bigskip 
      \begin{center}\begin{large}}{\end{large}\end{center} \end{quotation}}

\def\beq{\begin{equation}}
\def\eeq#1{\label{#1}\end{equation}}
\def\eeqn{\end{equation}}

\def\beqa{\begin{eqnarray}}
\def\eeqa#1{\label{#1}\end{eqnarray}}
\def\eeqan{\end{eqnarray}}

\let\bar=\overbar

\def\Dslash{\not{\hbox{\kern-4pt $D$}}}
\def\dslash{\not{\hbox{\kern-2pt $\del$}}}

\def\msb{{\bar{\ssstyle M \kern -1pt S}}}

\usepackage{color}

\newcommand\pubnumber{ CMS CR-2014/184 }

\newcommand\pubdate{\today}

\def\affiliation{
On behalf of the CMS Experiment, \\
Scuola Normale Superiore and INFN, Pisa }

\newcommand{\BsMuMu}{${B}^{0}_{s}\rightarrow \mu^+\mu^-\ $}
\newcommand{\BzsMuMu}{$ B_{(s)}^0 \rightarrow \mu^+\mu^-\ $}
\newcommand{\BMuMu}{${ B}^0 \rightarrow \mu^+\mu^-\ $}
\newcommand{\BRs}{${\mathcal B }( { B^{0}_{s}\rightarrow \mu^+\mu^-} ) $}
\newcommand{\BRz}{${\mathcal B }( { B^{0}\rightarrow \mu^+\mu^-} ) $}

\newcommand{\aBRs}{$\left<{\mathcal B }( { B^{0}_{s}\rightarrow \mu^+\mu^-} )\right> $}

\begin{document}

\large
\begin{titlepage}
\pubblock

\vfill
\Title{  $ B_{(s)}^0 \rightarrow \mu^+\mu^-\ $ at CMS  }
\vfill

\Author{ Franco Ligabue }
\Address{\affiliation}
\vfill
\begin{Abstract}The search for the rare \BMuMu and \BsMuMu decays in pp collisions at $\sqrt s = 7$~GeV and $\sqrt s = 8$~GeV, collected at the LHC in 2011 and 2012, is briefly reviewed. The data analyzed by CMS correspond to a total integrated luminosity of 5 and 20 $\rm fb^{-1}$, respectively. The time-integrated average branching fraction $\left<{\mathcal B }( { B^{0}_{s}\rightarrow \mu^+\mu^-} )\right>$ has been measured to be $(3.0 ^{+1.0}_{-0.9})\times 10^{-9}$ in accordance with the Standard Model predictions, while an upper limit $\left<{ \mathcal B }( B^{0}\rightarrow \mu^+\mu^-)\right> < 1.1\times 10^{-9}) $ has been placed on the other investigated decay at 95\% CL. A preliminary combination with the results from LHCb is also presented for both channels, and prospects for the future are briefly discussed.

\end{Abstract}
\vfill

\begin{Presented}
The Second Annual Conference\\
 on Large Hadron Collider Physics \\
Columbia University, New York, U.S.A \\ 
June 2-7, 2014
\end{Presented}
\vfill
\end{titlepage}
\def\thefootnote{\fnsymbol{footnote}}
\setcounter{footnote}{0}
%

\normalsize 


\section{Introduction}
The decays \BsMuMu and \BMuMu are strongly suppressed in the Standard Model. They are Flavour Changing Neutral Current processes that are forbidden at tree level and can only proceed through higher-order  (``box'' and ``penguin'') diagrams. Since they involve a scalar meson decaying weakly to two ultrarelativistic leptons, the decays are also helicity-suppressed. The purely leptonic final state 
makes it a relatively clean process to compute in the Standard Model, which predicts rather precise values for the branching ratios, namely $(3.23 \pm 0.27) \times 10^{-9}$ for ${\mathcal B }( {\rm B^{0}_{s}\rightarrow \mu^+\mu^-} ) $, and $(1.07 \pm 0.10) \times 10^{-10}$ for \BRz . These values are expected to be significantly altered in several  extensions of the Standard Model, which makes \BzsMuMu a particularly clean probe for New Physics~\cite{theory}. In some supersymmetric models, for instance, the sensitivity of \BRs\ to the fundamental parameter $\tan\beta$ can become as high as $\tan^6\beta$. The neutral-meson $\rm B_{(s)}^0$-$\rm \bar B_{(s)}^0$ system is also subject to flavour oscillation, which changes the observable time-integrated flavour average branching fraction with respect to the flavour-eigenstate decay probability computed at production time ($t=0$):

\begin{equation}
\left<{\mathcal{B}}( {\rm B^{0}_{s}\rightarrow \mu\mu} )\right>  = \frac {1+ {\cal A}^{\mu\mu}_{\Delta\Gamma}y_s}{1-y^2_s} \left.\mathcal B ( {\rm B^{0}_{s}\rightarrow \mu\mu} )  \right|_{t=0}
\end{equation}

where $y_s \equiv \frac {\Delta\Gamma}{2\Gamma}$ is the relative width difference between the two mass eigenstates, and where the parameter $A^{\mu\mu}_{\Delta\Gamma}$, which is unity in the Standard Model, can also be sensitive to New Physics. 

\section{CMS detector and muon identification}
The CMS detector has been described in detail in several publications (see for instance \cite{CMS_jinst}). Its relevant features here are basically its tracking, vertexing, and muon identification capabilities. The central silicon tracker, which is immersed in a 3.8~T uniform magnetic field parallel to the beam direction, can reconstruct charged tracks within a pseudo-rapidity range $|\eta | < 2.5$, with a $p_T$ resolution around 1\% and an impact parameter resolution of about $\rm 15~ \mu m$ in the kinematical range relevant to this analysis. Three different types of muon-specific detectors (drif tubes, cathode strip chambers and resistive plate chambers) are located in the return yoke of the magnetic field. 

Muons are very efficiently reconstructed and identified by combining silicon tracker and muon detector information. The reconstruction efficiency is checked on data with tag-and-probe techniques and is above 99\%  for a wide portion of the relevant phase space. In order to reduce the contamination from misidentified hadrons -- a crucial issue for this analysis -- a dedicated multivariate selection has been specially developed and applied on top of the standard CMS muon reconstruction. The variables used for this additional selection are either purely kinematic,  related to silicon tracker or muon detector information, or related to tracking fit quality. Adding this supplemental selection reduces the hadron misidentification rate by roughly 50\%, at the moderate expense of  a 10\% loss in efficiency. The final hadron misidentification probability ranges roughly from $0.5\times10^{-3}$ to $2.2\times10^{-3}$.

\section{Analysis strategy and data selection}

The measurement strategy for detecting \BzsMuMu decays in pp collisions is in principle straightforward: one must find a pair of oppositely charged muon tracks, compatible with originating from a common vertex, possibly displaced from the beam line, and whose invariant mass lies within a pre-defined interval centred around the $ B^0_{(s)}$ mass. The signal can then be extracted by event counting or by fitting the mass spectrum, after suitable background subtraction. In order to derive a branching fraction  from the measured yield, though, one needs to know very precisely the signal selection efficiency, as well as the $B^0_{(s)}$ production cross section and the absolute luminosity $\cal L$.
A way of reducing the large systematics involved is to refer the observed yield to that of a ``normalization'' channel whose branching ratio is rather precisely known, namely $B^\pm \rightarrow J / \Psi \,K^\pm \rightarrow (\mu^+\mu^-) K^\pm$,  by means of the following luminosity-independent formula:

\begin{equation}\label{eq:BRextr}
\left<{\cal B} (B_s^0\rightarrow\mu^+\mu^- )\right> = 
\frac {N_{\rm obs}^{B^0_s}}{N_{\rm obs}^{B^\pm}}
\times \frac{\varepsilon_{B^\pm}}{\varepsilon_{B_s^0}} 
\times \frac{f_u}{f_s}
\times {\cal B} (B^\pm \rightarrow J/\Psi\,K^\pm ) 
\times {\cal B} (J/\Psi \rightarrow\mu^+\mu^- )
\end{equation}

where the first factor in the multiplication is the ratio of the observed number of events for the signal and for the normalization channel, and the second is the ratio of detection efficiency, where many systematics uncertainties cancel to first order. The third factor ($f_u/f_s$) is the production ratio of $B^\pm$ to $B^0_s$ mesons at LHC, and is taken from experiment~\cite{LHCbfufs}.

The analyzed pp collision samples correspond to an integrated luminosity of 
$\rm 5 \,fb^{-1}$ 
collected by CMS in 2011 at $\sqrt{s} = \rm 7 TeV$, 
and $\rm 20 \, fb^{-1}$ collected in 2012 at $\sqrt{s} = 8 \rm TeV$.
Events are selected  via a two-level (online and offline) trigger, requiring the presence of two muons with an invariant mass within a specified window and with a minimum common vertex  fit probability. 
After applying the multivariate selection for enhancing the muon purity, two adjacent signal  regions and two sideband windows are defined within the overall accepted invariant mass region between 4.9 and 5.9 $\rm GeV/c^2$. The data  falling in the signal region, between 5.20 and 5.45~$\rm GeV/c^2$ are then never looked at (``blind'' analysis) except in the final stages of signal yield extraction.

Exclusively reconstructed $B^\pm \rightarrow J/\Psi\,K^\pm$ and $B_s^0 \rightarrow (J/\Psi)\phi \rightarrow (\mu^+\mu^-)(K^+K^-)$, selected with kinematic cuts similar to those used for the signal, provide the so-called ``normalization'' and ``control'' samples. The former is used for branching ratio extraction according to eq.~\ref{eq:BRextr}, both --  but especially the latter -- are used for data-simulation comparison and cross checks.

\begin{figure}[htb]
\centering
\includegraphics[height=2in]{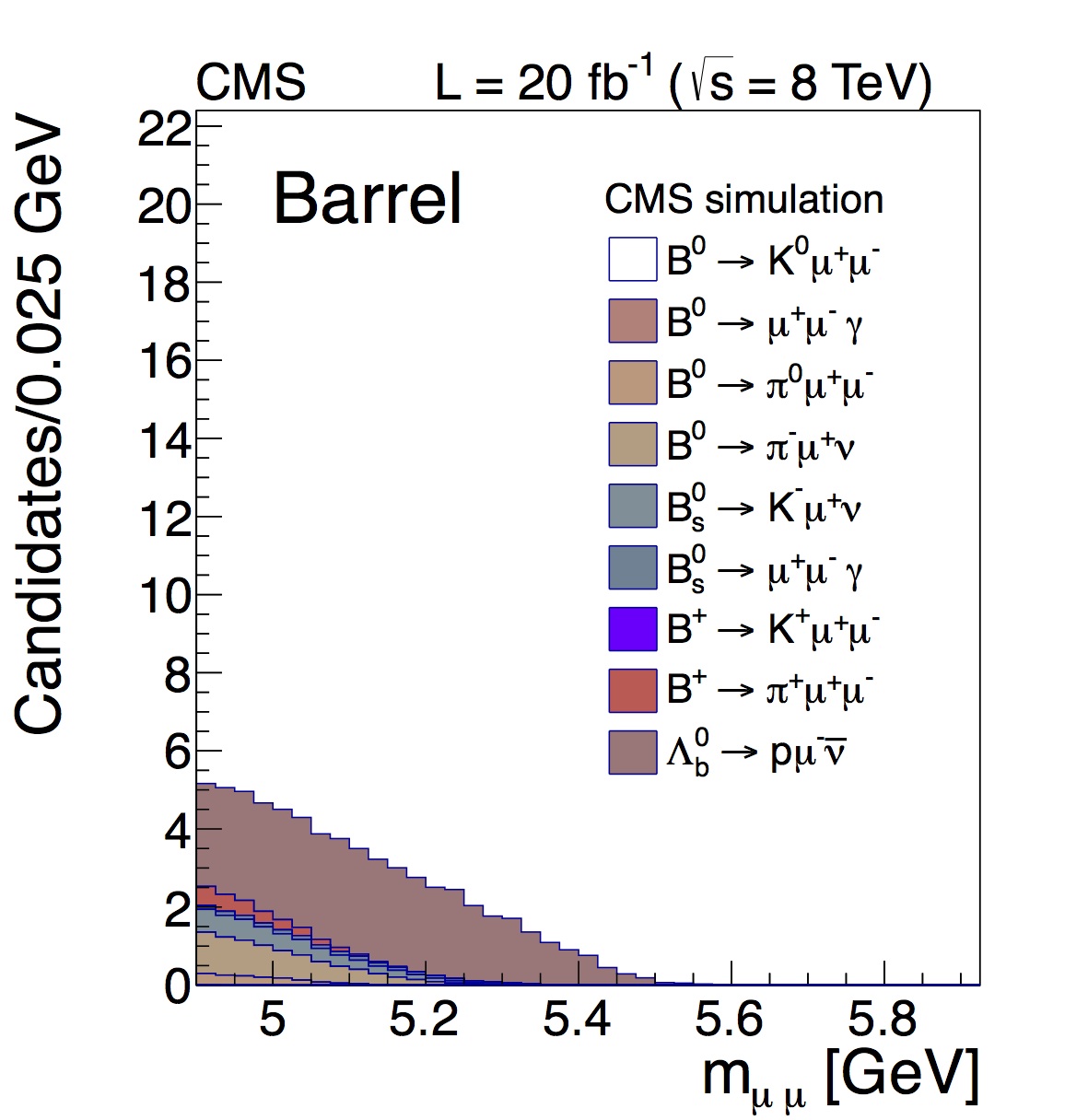}
\includegraphics[height=2in]{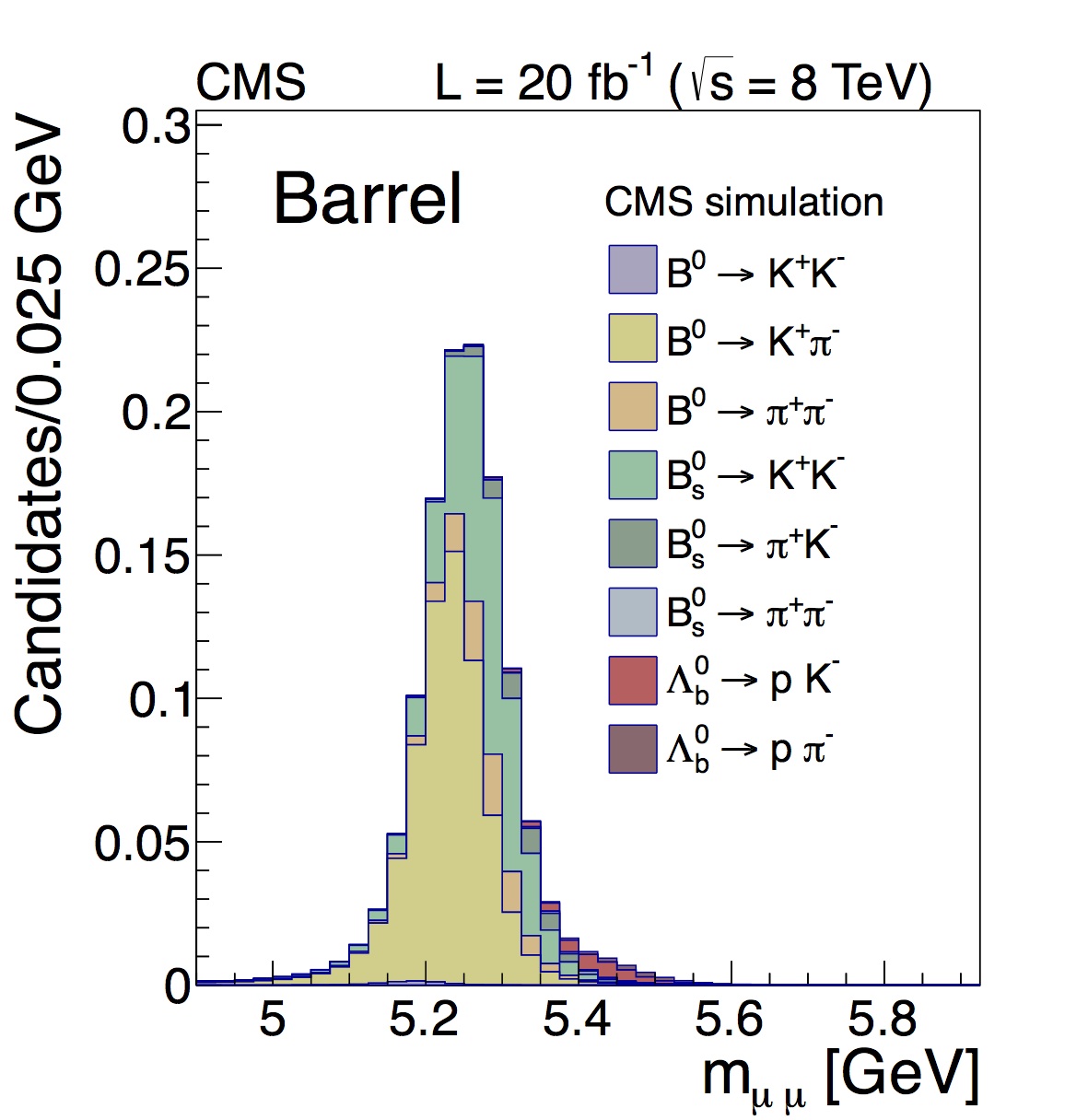}
\caption{Dimuon invariant mass distribution for background events from rare 3-body semileptonic (left) and 2-body hadronic decays single b-hadrons in simulation.}
\label{fig:bkg}
\end{figure}

Most of the background events mimicking the signal involve the actual presence of $b$ hadrons. Events where two real muons coming from semileptonic decays of separate $b$ hadrons appear by chance to  form the right mass (``combinatorial background'') are more or less flatly distributed in the invariant mass window, whereas events where the two selected tracks (either two real muons or a real muon and a misidentified hadron) come from a three-body semileptonic decay of a single b-hadron (``semileptonic background'') exhibit a monotonically sloping distribution (see left plot in fig.~\ref{fig:bkg}).
Events where both the decay products of two-body b-hadrons decays (pions, kaons or protons) have been misidentified as muons are much more dangerous since they peak at the B mass (``peaking background'', see right plot in fig.~\ref{fig:bkg}) and their contribution cannot be interpolated from the data sidebands like for the previous category.

\begin{figure}[htb]
\centering
\includegraphics[height=2in]{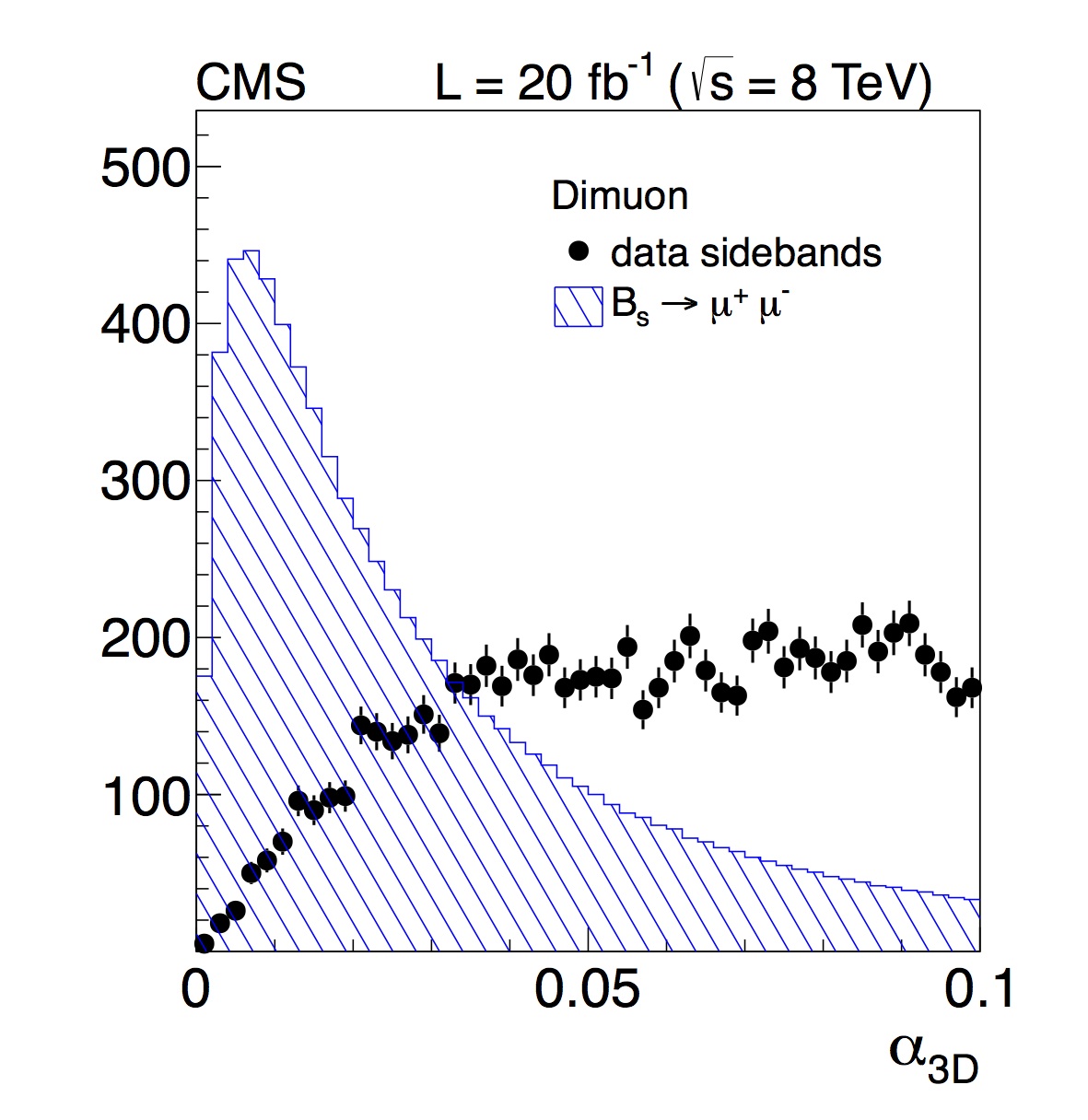}
\includegraphics[height=2in]{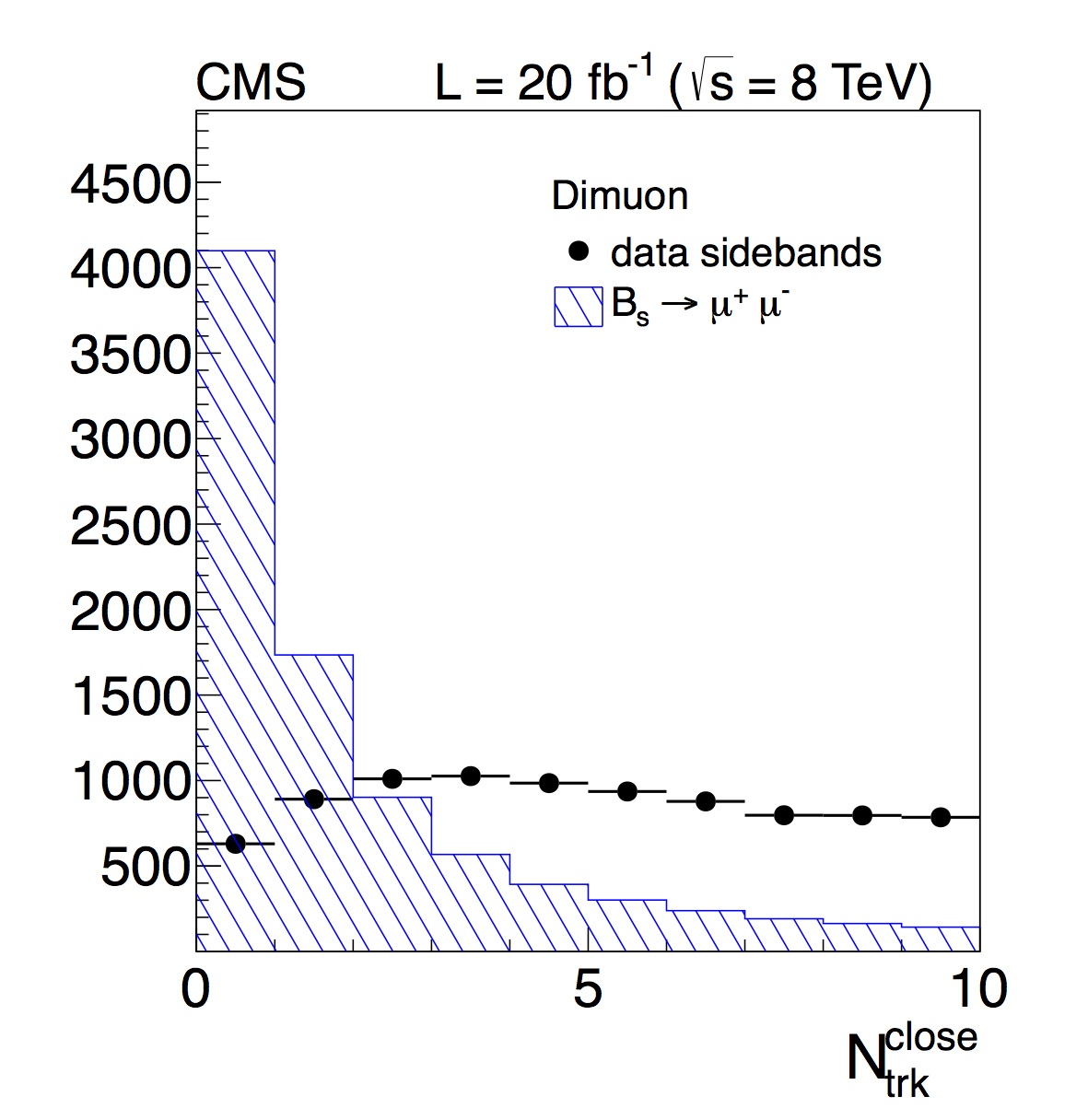}
\caption{Distribution of two of the most effective variables used in the BDT technique, for simulated signal events (histogram) and data non-signal events from mass sidebands (data points). { Left}: pointing angle between B candidate momentum and primary-to-secondary vertex direction. { Right}: number of tracks in the vicinity of the B candidate.}\label{fig:var}
\end{figure}

A multivariate  (Boosted Decision Tree or BDT) technique~\cite{TMVA} is used to build a discriminating variable providing a good separation of the signal from the combinatorial background. The most effective  variables used in the combination turn out to be either related to the secondary vertex separation (the impact parameter and its significance, or the  angle between the dimuon momentum  and the vector pointing from the primary to the secondary vertex), or else ``isolation'' variables,  such as the number of tracks close to the dimuon system, shown in the right plot of fig.~\ref{fig:var}.
The BDT has been optimized using simulated events, as representative of the signal, and real data from the mass sideband windows, as representative of the combinatorial background.


\section{Signal extraction and results}

The events selected in the \BMuMu signal mass window have been analyzed with a simple counting experiment after a cut on the value of the BDT discriminating variable optimizing the sensitivity $S/\sqrt{S+B}$. The events were split into four samples according to the year of data taking and the pseudorapidity of the detected muons (``barrel'' and ``endcap'' events). The signal select efficiency is evaluated to range roughly from 1 to 3 permille.

\begin{figure}[htb]
\centering
\includegraphics[height=2in]{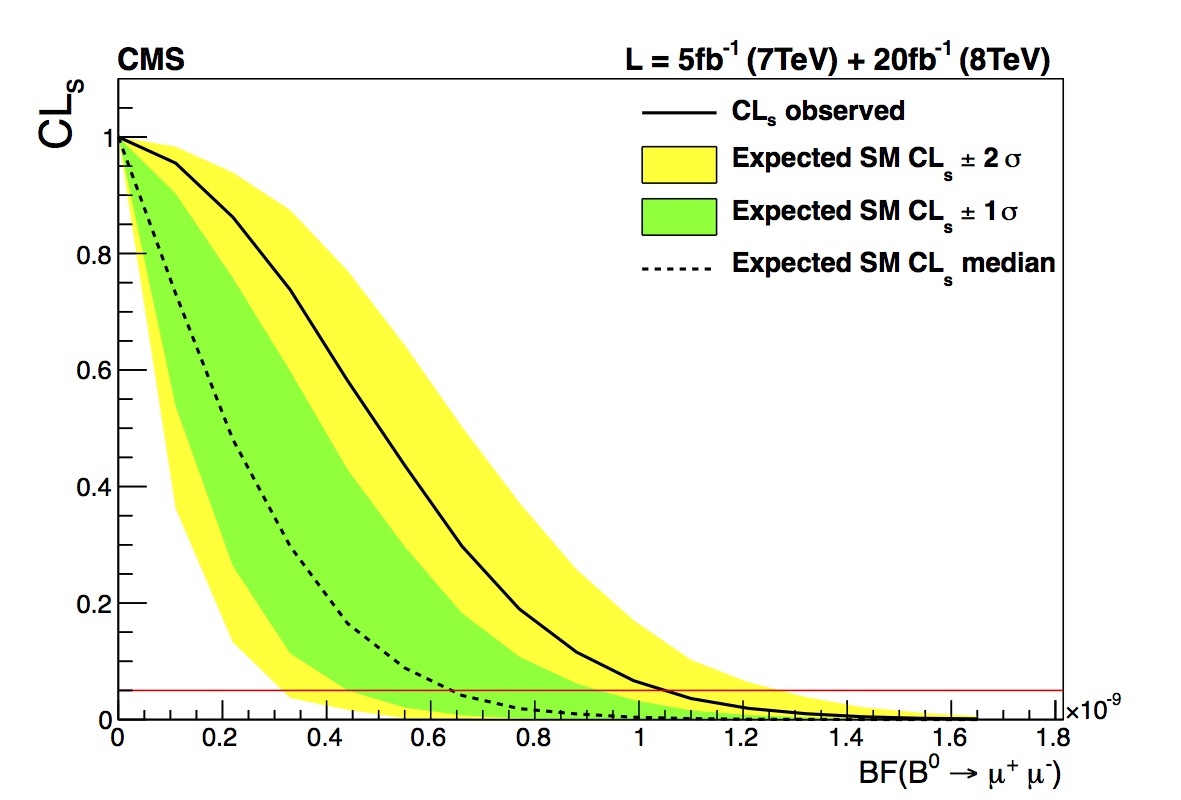}
\includegraphics[height=2in]{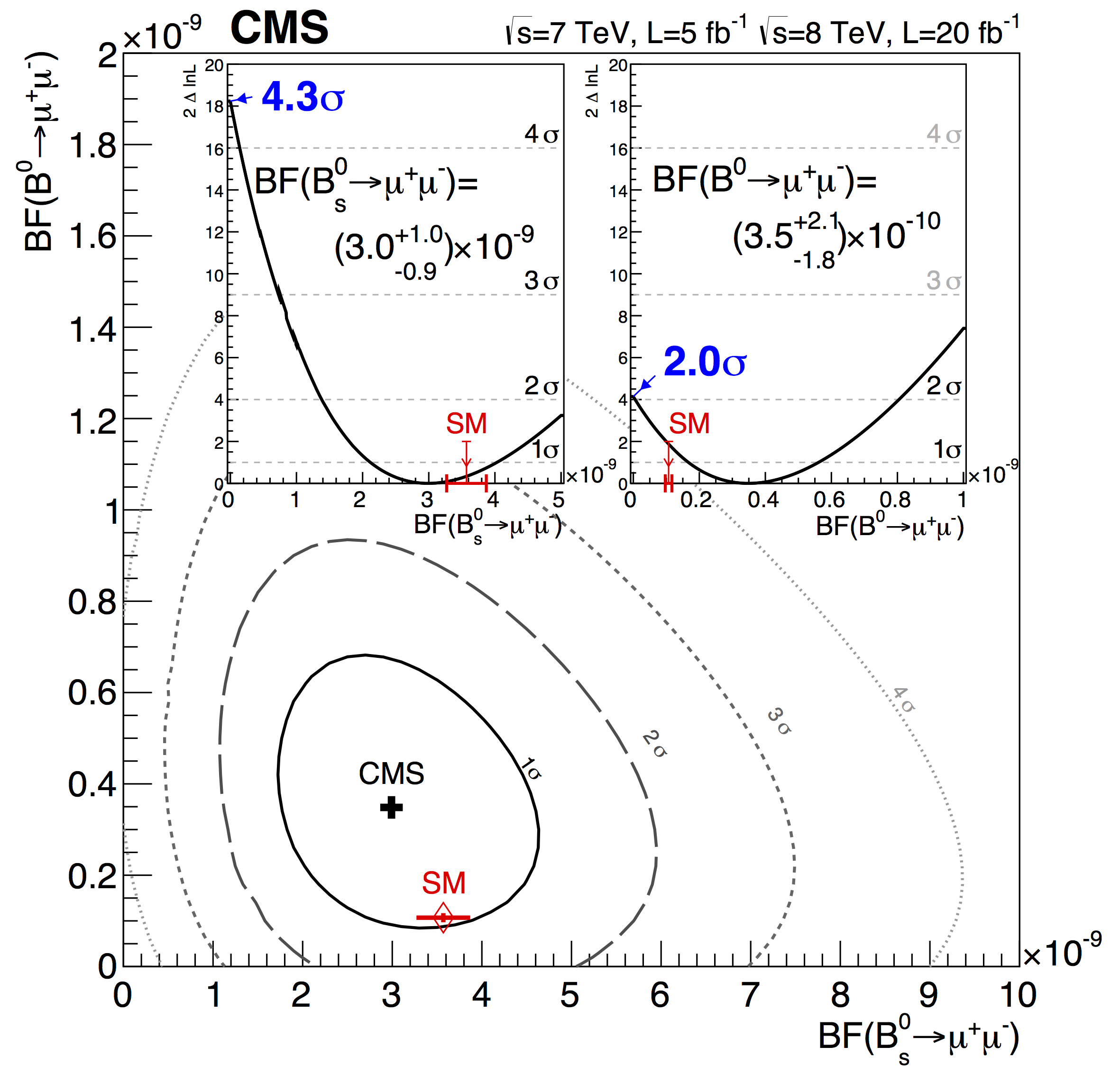}
\caption{Left: Observed and expected 95\% CL limit as a function of the \BMuMu assumed branching fraction.
Right: Scan of the joint likelihood ratio for the two branching fractions. As insets, the likelihood ratio scans for each of the branching fraction when the other is taken as a nuisance parameter.}
\label{fig:results}
\end{figure}

No significant excess is observed for this channel, and an upper limit is placed on the branching ratio at 95\% confidence level using the $CL_s$ approach, obtaining

\[
\left<{\cal B} (B^0\to\mu^+\mu^-)\right>  < 1.1 \times 10^{-9}
\]
which is compatible with the Standard Model predicted limit at the $2\sigma$ level, though on the higher side, as shown in the left plot of fig.~\ref{fig:results}.

\begin{figure}[htb]
\centering
\includegraphics[height=2in]{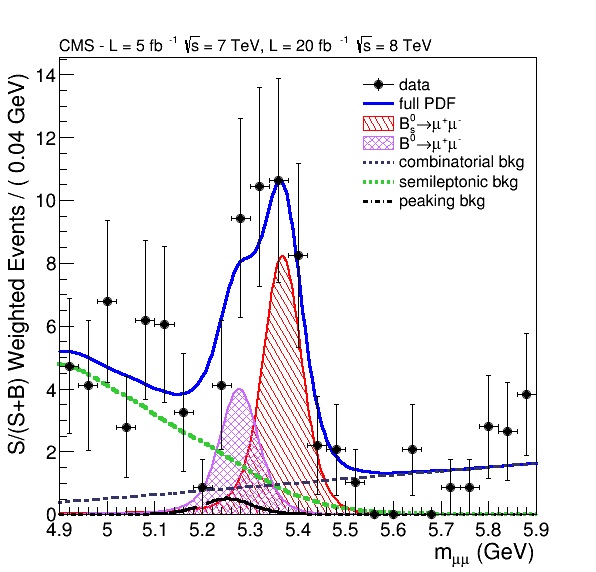}
\caption{Plot illustrating the weighted average of the fit results on the invariant mass spectrum in the 12 data categories, for  \BsMuMu yield extraction. The weights are proportional to the signal purity $S/(S+B)$ determined at the $B^0_s$ peak position.}
\label{fig:data_weighted}
\end{figure}

The \BsMuMu signal is extracted with an unbinned maximum likelihood fit to the invariant mass spectrum applied after splitting the data into twelve subsamples or ``categories'' according to the value of the BDT discriminating variable and, again, the year of data taking and the pseudorapidity of the detected muons. The size and boundaries the BDT intervals have been chosen so as to have roughly the same number of expected signal events in each category. For each category the signal shape is described by a Crystal Ball function with a  variable (per-event) mass-resolution parameter. The combinatorial background is described by a straight line, while the shapes of the bakgrounds from rare b-hadron decays are taken from simulation. The estimated systematic uncertainties (for instance due to uncertainties on hadron misidentification probability, or on the peaking background normalization)
are included in the fit as further Gaussian constraints. 
A signal excess is observed, as is visible in the illustrative plot of fig.~\ref{fig:data_weighted}, corresponding to a measured value for the time-integrated branching fraction of
\[
\left<{\cal B} (B^0_s\to\mu^+\mu^-) \right>= (3.0^{+1.0}_{-0.9}) \times 10^{-9}
\]
where the quoted uncertainty includes both the (dominating) statistical component and the systematic contributions. The systematics is dominated by the uncertainty on the value of the $f_u/f_s$ factor in eq.~\ref{eq:BRextr}, which is taken from a published LHCb measurement~\cite{LHCbfufs}.

The right plot in fig.~\ref{fig:results} shows the contours of the joint likelihood for the two branching ratios, along with the Standard Model prediction, which appears to be compatible at the 1$\sigma$ level with the results. The top right inset in the same plot shows that the observed \BMuMu excess is at the level of $2.0\sigma$ with respect to the background-only hypothesis.

\section {Combination with LHCb and prospects}

The CMS measurement followed analogous results published by the LHCb collaboration~\cite{LHCbpaper}. While a proper combination, which requires building a joint likelihood, is being worked at, preliminary combined results have been obtained by the two collaborations for both branching fractions, by simply taking into account correlated and uncorrelated uncertainties:

\[
\left<{\cal B} (B^0_s\to\mu^+\mu^-)\right> = (2.9\pm 0.7) \times 10^{-9}
\]

\[
\left<{\cal B} (B^0\to\mu^+\mu^-)\right>  = (3.6^{+1.6}_{-1.4}) \times 10^{-10}
\]

The dominating correlated uncertainty is of course the value of $f_u/f_s$, which is the same for the two experiments. Although not final, the combined value for \aBRs\ has a significance above $5.0\, \sigma$. Both values are compatible with the SM expectation, which can help to reduce considerably the allowed parameter space in many New Physics scenarios.

The future prospects for the measurements have been investigated by CMS~\cite{FTR}, also in view of the foreseen (tracking) detector upgrade. The future LHC runs will be characterized by higher  centre-of-mass energy, bunch-crossing frequency, and  luminosity. This will lead to an increase of the signal production, partly due to the increased cross-section, which will have to meet with several serious challenges, especially concerning trigger rates and high pile-up level (which might affect isolation and therefore potentially decrease the selection efficiency). 
Improvements in the analysis are being studied, particularly as far as fake muon rejection is concerned (crucial for reducing the peaking background). 

By rescaling the present results to the future foreseen luminosities and centre-of-mass energies, assuming mostly unchanged selection efficiencies, the significance of the \BsMuMu branching fraction measurement is expected to double by the end of 2017. The significance of the \BMuMu measurement is expected to reach $5\, \sigma$ by the end of the HL-LHC phase.


\end{document}